\documentstyle[11pt,twoside]{article}
\begin{document}

\def\thefootnote{\fnsymbol{footnote}}

\begin{center}
{\large\bf Optical Gravitational Lensing Experiment.\\
Photometry of the MACHO-SMC-1 Microlensing Candidate.
\footnote{Based on observations obtained with the 1.3~m Warsaw telescope
at the Las Campanas Observatory of the Carnegie Institution of
Washington.}}
\vskip1.3cm
{\large A.~~U~d~a~l~s~k~i$^1$,~~M.~~S~z~y~m~a~{\'n}~s~k~i$^1$,~~M.~~K~u~b~i~a~k$^1$,\\ 
G.~~P~i~e~t~r~z~y~\'n~s~k~i$^1$,~~ 
P.~~W~o~\'z~n~i~a~k$^2$,~~ and~~K.~~\.Z~e~b~r~u~\'n$^1$}
\vskip5mm
{$^1$Warsaw University Observatory, Al.~Ujazdowskie~4, 00-478~Warszawa, Poland\\
e-mail: (udalski,msz,mk,pietrzyn,zebrun)@sirius.astrouw.edu.pl\\
$^2$ Princeton University Observatory, Princeton, NJ 08544-1001, USA\\
e-mail: wozniak@astro.princeton.edu}
\end{center}

\vskip13mm

{\footnotesize\begin{center}{ABSTRACT}\end{center}
We present photometric observations of the MACHO-SMC-1
microlensing  candidate collected by the OGLE-2 project. We show light
curves of both  components of the 1.6 arcsec blend: microlensed star and
its optical companion. We find the  contribution of the optical
companion to the total  flux to be 24\% and confirm presence of the
small amplitude periodic  oscillations in the light curve of the lensed
star with the period of 5.096 days and amplitude 0.05~mag.
The lensed star is probably an ellipsoidal binary system.}

\vskip1cm
{\bf 1. Introduction}
\vspace*{9pt}

Discovery of the first microlensing event candidate toward the Small 
Magellanic Cloud by the MACHO Collaboration (Alcock {\it et al.} 1997) opens a new 
direction in which probing of the Galactic halo with microlensing phenomena 
can be possible. Although it may take years before a significant sample of 
events in this direction will be collected allowing to draw sound conclusions 
about distribution of mass in the Galactic halo, the first candidate itself 
seems to be very interesting. 

The microlensed source is a relatively bright (${V\sim 17.7}$~mag) main
sequence  star blended with a fainter object. Contribution of the
companion was  estimated to be 23--28\% of the total blend light. The
event duration was  242 days (Einstein diameter crossing time) and the 
magnification reached maximum of 2.4 on  January 11, 1997  (Alcock {\it et 
al.} 1997; parameters derived taking into account blending). Symmetric light 
curve and good  achromaticity make this brightening an excellent microlensing 
event candidate. However, it should be noted that the position of the lensed 
star on the color-magnitude diagram is dangerously close to the non-periodic 
variable  stars which can  mimic microlensing. 

The MACHO-SMC-1 candidate was also observed by the EROS group 
(Palan\-que-Delabrouille {\it et al.} 1997). They covered mostly the rising branch of 
the event. They found that the lensed star exhibits small, periodic variations 
of light with the amplitude of a few percent and the period of 5.123 days. 
However, due to poor resolution they were only able to perform photometry of 
the entire blend and could not  assign those possible variations to any of the 
blended components. 

Parameters of the event, in particular its long duration, suggest that the 
lensing object could be a massive body (${\sim2.5{\rm M}_{\odot}}$) in the Galactic 
halo or alternatively an object in the SMC (Alcock {\it et al.} 1997). 

The SMC is one of the targets of the second phase of the Optical
Gravitational  Lensing Experiment -- OGLE-2 (Udalski, Kubiak and
Szyma\'nski 1997). Although  observations of the majority of fields in
the SMC started in June 1997, well  after the event reached maximum of
light, we decided to analyze OGLE-2 data of  the MACHO-SMC-1 event to
clear up the questions concerning possible  variability of the lensed
star  and contribution of the blend components to  the total flux. In
this paper we present our results.

\vspace*{12pt}
{\bf 2. Observations}
\vspace*{9pt}

The OGLE-2 project observations are carried out at the Las Campanas 
Observatory, Chile, which is operated by the Carnegie Institution of 
Washington, with the 1.3-m Warsaw telescope equipped with the "first 
generation" CCD camera. Details of the equipment and data reduction techniques 
can be found in Udalski, Kubiak and Szyma\'nski (1997). 

Observations of the SMC are made in the driftscan mode with drifts in 
declination. One scan covers approximately ${14\times56}$ arcmins on the sky. 
Majority of scans are obtained with the $I$-band filter, with some 
measurements in $V$ and $B$-bands. The effective exposure time is 120~sec, 
180~sec and 240~sec for $I$, $V$, and $B$-bands, respectively. Observations of 
the field in which the MACHO-SMC-1 candidate was identified -- SMC\_SC8 -- 
started on Jun.~27, 1997. The data presented in this paper cover the period 
through Oct.~9, 1997 (JD~2450730.6). 

\vspace*{12pt}
{\bf 3. Results} 
\vspace*{9pt}

Fig.~1 shows a ${30\times30}$ arcsec subframe centered on the
MACHO-SMC-1  candidate taken at 1.0 arcsec seeing conditions. As can be
seen the blend is  easily resolved and both components could be easily
measured independently  with the OGLE-2 software. The $I$-band light
curves of both stars are  presented in Figs.~2 and 3. The zero point was
chosen as the normal  (off-event) brightness of the lensed star,
calculated by fitting to  our observations the  theoretical microlensing
light curve with Alcock {\it et al.} (1997) parameters. Error bars correspond
to the errors returned by the  photometry  program (DoPhot) rescaled by
a factor of 1.3 to approximate the  observational errors. Errors of both
components, in particular the fainter  component of the blend, are
larger than for typical stars of such brightness  due to relatively
small separation. The distance between components is 1.6  arcsec. 

As can be seen from Figs.~2 and 3, the magnified star was the brighter 
component of the blend. Although the observations begun more than 5
months  after the maximum of brightness, the star was still slightly
above its normal  brightness and the slow fading can still be noticed.
The fainter component of  the blend was constant over the entire period
of observations. The normal  state brightness of the lensed star  and
the mean magnitude of the fainter companion suggest that, in the
$I$-band, the brighter component contributes  76\% to the total 
light. Both components of the blend have almost  exactly the same
${V-I}$ colors. Thus, in the $V$-band contribution of the fainter
component is very  similar. Both stars are located among the SMC main
sequence stars on the color magnitude diagram.

To check for periodic variability reported by Palanque-Delabrouille
{\it et al.}  (1997) we performed a period analysis  of the light curves of
both components with  the {\sc clean} algorithm (Roberts, Leh\'ar and
Dreher 1987). For the brighter star we first rectified the light  curve
by subtracting microlensing brightening. The analysis yields the  
periodicity of 5.096~days for the brighter star. No significant period
was  found for the fainter star. We also checked photometry of two
nearby stars of similar brightness and again no  significant periodicity
was found. 

Fig.~4 shows rectified observations of the brighter star folded with the 
period of 5.096~day. Clear sinusoidal variations can be noticed. We 
fit a sinusoid to the data. The full amplitude (peak to peak) of the best fit  
is about 0.05~mag. Elements of minimum of brightness are given by the 
following equation: 
$$\begin{tabular}{r@{\hspace{3pt}}c@{\hspace{3pt}}c@{\hspace{3pt}}r@{\hspace{3pt}}c@{\hspace{3pt}}l}
{\rm JDhel.} & = &        & 2450625.612 & +     & 5.096$\times P$\\
             &   & $\pm$  &       0.050 & $\pm$ & 0.025\,.
\end{tabular}
$$

\vspace*{12pt}
{\bf 4. Conclusions}
\vspace*{9pt}

Analysis of the photometric data of the MACHO-SMC-1 microlensing event 
candidate collected in the course of the OGLE-2 program indicates that
the  star which underwent magnification was the brighter component of
the blend separated by 1.6  arcsec. Light curve of the fainter component
shows no  significant light variations. The brighter star contributes
76\% to the blend  light in the $V$ and $I$-bands. 

The light curve of the microlensing candidate shows additional small
amplitude  periodic variations with the period of 5.096~days -- close
to the value  reported by the EROS group. Also the amplitude of
sinusoidal variations is  similar. As our data cover mostly the
off-microlensing light curve while the  EROS data were taken during the
event, this may suggest that the amplitude is  constant and is not
related to microlensing. Most likely the star is a  binary system with
one or both components ellipsoidally distorted. Changing  aspects cause
small amplitude, sinusoidal variations similar to those observed  in the
light curve of the lensed star. The real period of the binary system 
would be twice of that derived in Section~3, that is 10.19~days. As the
star  is relatively bright, this hypothesis can be relatively easy 
verified by spectroscopic observations. 

Photometry of the  MACHO-SMC-1 candidate can be retrieved from the OGLE network 
archive: {\it ftp://sirius.astrouw.edu.pl/ogle/ogle2/macho-smc-1.} 

\newcommand{\Acknow}[1]{\par\vspace{5mm}{\bf Acknowledgements.} #1}

\Acknow{We thank Prof.\ B.~Paczy\'nski for his comments and
and Dr.\ I.~Semeniuk for double checking period analysis.
This paper was partly supported from the KBN BST grant. Partial  support
for the OGLE  project was provided with the NSF grant AST-9530478 to
B.~Paczy\'nski.} 

\newenvironment{references}%
{
\footnotesize \frenchspacing
\renewcommand{\thesection}{}
\renewcommand{\in}{{\rm in }}
\renewcommand{\AA}{Astron.\ Astrophys.}
\newcommand{\AAS}{Astron.~Astrophys.~Suppl.~Ser.}
\newcommand{\ApJ}{Astrophys.\ J.}
\newcommand{\ApJS}{Astrophys.\ J.~Suppl.~Ser.}
\newcommand{\ApJL}{Astrophys.\ J.~Letters}
\newcommand{\AJ}{Astron.\ J.}
\newcommand{\IBVS}{IBVS}
\newcommand{\PASP}{P.A.S.P.}
\newcommand{\Acta}{Acta Astron.}
\newcommand{\MNRAS}{MNRAS}
\renewcommand{\and}{{\rm and }}
\section{{\rm REFERENCES}}
\sloppy \hyphenpenalty10000
\begin{list}{}{\leftmargin1cm\listparindent-1cm
\itemindent\listparindent\parsep0pt\itemsep0pt}}%
{\end{list}\vspace{2mm}}

\def\TYLDA{~}
\newlength{\DW}
\settowidth{\DW}{0}
\newcommand{\dw}{\hspace{\DW}}

\newcommand{\refitem}[5]{\item[]{#1} #2%
\def\REFARG{#3}\ifx\REFARG\TYLDA\else, {\it#3}\fi
\def\REFARG{#4}\ifx\REFARG\TYLDA\else, {\bf#4}\fi
\def\REFARG{#5}\ifx\REFARG\TYLDA\else, {#5}\fi.}

\vskip3cm

\begin{center}
{\bf Figure Captions}
\end{center}

Fig.~1. ${30\times30}$~arcsec subframe centered on the MACHO-SMC-1
microlensing event candidate. North is up and East to the left.

Fig.~2. The $I$-band light curve of the microlensed star. Solid line shows the 
theoretical microlensing light curve of MACHO-SMC-1. Broken line indicates the 
normal brightness of the star.

Fig.~3. The $I$-band light curve of the star separated by 1.6 arcsec
from the microlensed star. Broken line indicates its mean magnitude.

Fig.~4. Observations of the microlensed star phased with the period of
5.096 days. Thick, solid line indicates the best fit sinusoid. Two
cycles are repeated for clarity.
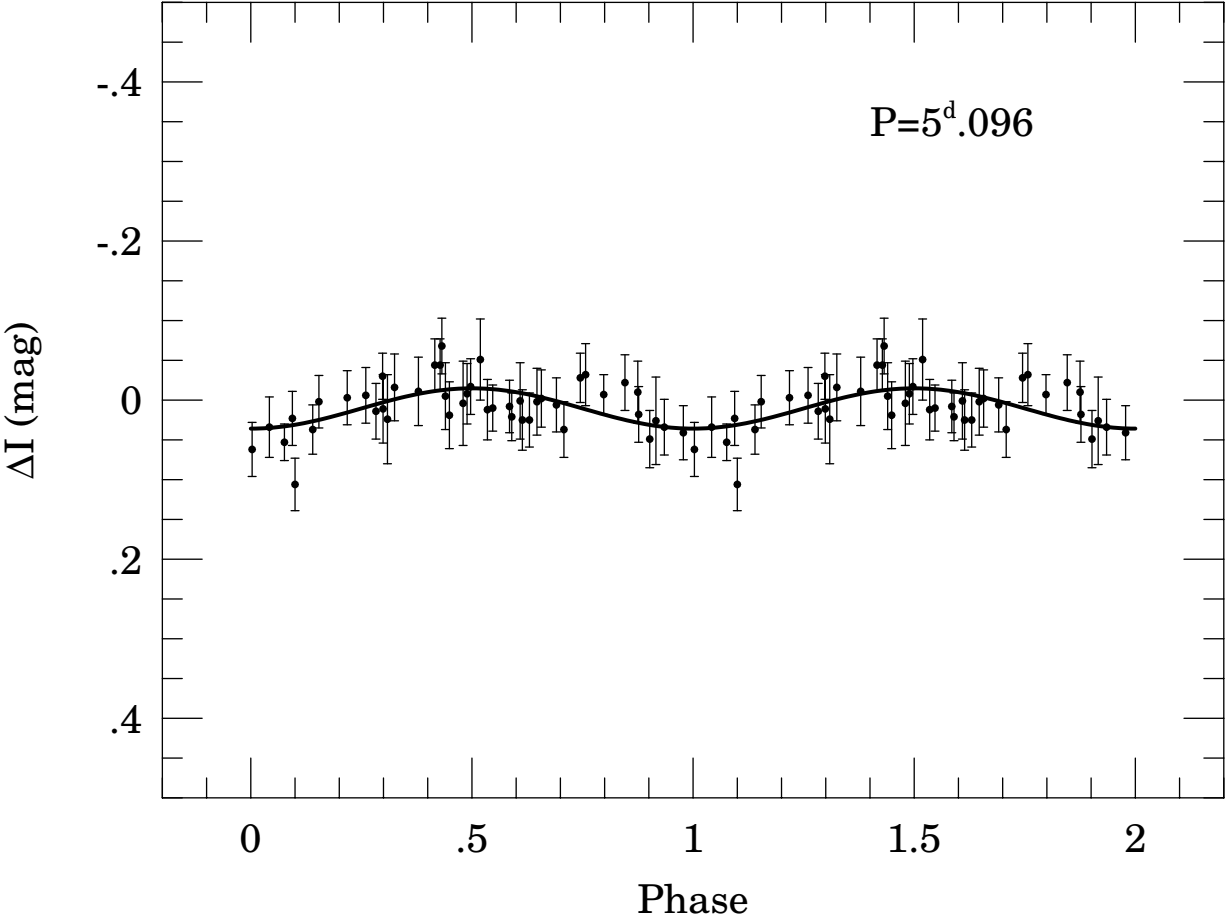
\end{document}